\documentclass[11pt]{article}
\usepackage{blois,epsfig}

\def\be{\begin{equation}}
\def\ee{\end{equation}}
\def\bea{\begin{eqnarray}}
\def\eea{\end{eqnarray}}

\def\lcdm{$\Lambda$CDM~}

\def\ltsima{$\; \buildrel < \over \sim \;$}
\def\gtsima{$\; \buildrel > \over \sim \;$}
\def\simlt{\lower.5ex\hbox{\ltsima}}
\def\simgt{\lower.5ex\hbox{\gtsima}}

\def\q{{\hat n} }

\begin{document}
\vspace*{4cm}
\title{Cross-correlating the Microwave Sky with Galaxy Surveys}

\author{Pablo Fosalba$^{1,2}$, Enrique Gazta\~{n}aga$^{3,4}$, Francisco J
Castander$^{3}$}

\address{
$1$ Institut d'Astrophysique de Paris, 98bis Bd Arago,
75014 Paris, France \\ 
$2$ Institute for Astronomy, University of Hawaii, 2680, Woodlawn Drive,
HI-96822, Honolulu, USA \\
$3$ Institut d'Estudis Espacials de Catalunya/CSIC, Gran
Capit\`a 2-4, 08034 Barcelona, Spain \\
$4$ INAOE, Astrofisica, Tonantzintla, Puebla 7200, Mexico}

\maketitle
\abstracts{
We present results for the cross-correlation 
between the WMAP 1st-year cosmic microwave background (CMB) anisotropy data 
and optical galaxy surveys: the APM and SDSS DR1 catalogs. 
Our measurement of a positive CMB-galaxy correlation 
on large angles ($\theta>4^\circ$)
yields significant detections of the
Integrated Sachs-Wolfe (ISW) effect and provides a new 
estimate of dark-energy in the universe, $\Omega_\Lambda=0.69-0.86$ (2$\sigma$ range). In addition, the correlated signal on small angles ($\theta<1^\circ$)  
reveals the imprint left by hot intra-cluster gas in the CMB photons:
the thermal Sunyaev-Zeldovich (SZ) effect.}

\section{Introduction}
The pattern of primary temperature anisotropies of the
cosmic microwave background (CMB) is expected to be distorted by
the large-scale structures of the universe as microwave photons 
travel from the last scattering surface (z $\simeq$ 1100) to us 
(see e.g \cite{hu/dodelson:2002}).
On large-scales (i.e, angular scales comparable to large clusters 
and super-clusters as seen in projection) , 
the observationally favored flat \lcdm models
predict such distortion is mainly produced by the 
energy injection photons experience
as they cross time-evolving dark-matter gravitational potential wells:
the so-called integrated Sachs-Wolfe effect (ISW) 
\cite{sachs/wolfe:1967,kofman/aas:1985}. On smaller scales,  
the primary CMB anisotropy pattern is altered when primordial photons
scatter off free electrons in the hot intra-cluster gas: the thermal 
Sunyaev-Zeldovich effect (SZ) \cite{sunyaev/zel:1980}.
 
Although these secondary anisotropies can be potentially measured from CMB maps
alone, in practice the ISW detection is severely limited by primary CMB 
anisotropies and cosmic variance, whereas the SZ requires high-spatial
resolution, multi-frequency observations and it is potentially 
contaminated by point sources.
Alternatively, large-scale structure tracers, such as galaxy surveys, 
provide a unique window to probe the
baryon and dark-matter distribution at intermediate redshifts (z$\simlt 2$) 
and its imprint in microwave photons
without significant confusion from other cosmological signals 
\cite{crittenden/turok:1996,peiris/spergel:2000,refregier/etal:2000}.
The combination of nearly full-sky high-sensitivity CMB maps 
obtained by WMAP \cite{bennett/etal:2003} 
with wide large-scale structure surveys 
has recently led to the first detections
of the ISW effect, setting new constraints on the dark-energy
content of the universe \cite{boughn/crittenden:2003,nolta/etal:2003}. 
It has also allowed probing the SZ effect with cluster templates 
\cite{diego/etal:2003,hernandez/rubino:2003,myers:2003}.

Here we concentrate on the cross-correlation of the cosmic
CMB anisotropies measured by WMAP
with galaxy number count fluctuations in the APM 
Survey \cite{maddox/etal:1990} and the first data release of the
Sloan Digital Sky Survey 
\footnote{http://www.sdss.org/dr1} \cite{abazajian/etal:2003} (SDSS DR1).
Our CMB-galaxy correlation analyses find significant detections for
both the integrated Sachs-Wolfe (ISW) and thermal Sunyaev-Zeldovich
(SZ) effects. 
 The reported ISW detection is in good agreement with previous
analyses based on X-ray and radio sources 
\cite{boughn/crittenden:2003,nolta/etal:2003}.
Further details on the results presented here
are given in \cite{fosalba/gazta:2003,fosalba/gazta/fjc:2003}. 
A similar analysis using SDSS data was presented in \cite{scranton/etal:2003}
while, more recently, \cite{afshordi/etal:2003} 
computed the CMB-galaxy correlation for 2MASS galaxies.

\section{Data}
\label{sec:data}

We make use of the largest datasets currently available to study the
CMB-galaxy cross-correlation. In order to probe the galaxy
distribution, we have used the APM survey as well as 
selected subsamples from the SDSS DR1.

The APM Galaxy Survey
is based on 185 UK IIIA-J Schmidt photographic plates each corresponding
to $5.8\times 5.8$ deg$^2$ on the sky limited to $b_J \simeq 20.5$ and
having a mean depth of $\simeq 400$ Mpc/h for $b <-40$ deg and $\delta<-20$
deg.  These fields where scanned by the APM machine and carefully matched
using the $5.8\times 0.8$ deg$^2$ plate overlaps. 
Out of the APM Survey we considered a $17<b_J<20$
magnitude slice, which includes 1.2 million galaxies at a mean redshift
$\bar{z} = 0.15$, in an equal-area projection pixel map with a
resolution of $3.5'$, that covers over $4300$ deg$^2$
around the SGC. 

The SDSS DR1 survey covers $\sim
2000$ deg$^2$ (i.e, 5 $\%$ of the sky).  The samples we analyzed 
have different redshift distributions and a large number of galaxies
(10$^5$-10$^6$, depending on the sample). We concentrate our analysis
on the North sky ($\sim$ 1500 deg$^2$, ie, 3.6 $\%$ of the sky),
because it contains the largest and wider strips. The South SDSS DR1
($\sim$ 500 deg$^2$) consists of 3 narrow and disjoint $2.5^\circ$
strips, which are less adequate for our analysis.
Our main sample, hereafter {\it SDSS all}, includes all objects
classified as galaxies with extinction corrected magnitude $r < 21$,
and a low associated error ($< 20 \%$). This sample contains $\sim$ 5
million galaxies distributed over the North sky. Its
predicted redshift distribution is broad and has a median redshift 
$\overline{z}\sim 0.3$. Our color selected high-redshift sample 
({\it SDSS high-z} thereafter)
comprises $\sim 3 \times 10^5$ galaxies, with $\overline{z}\sim 0.5$.

For the CMB data, we use the first-year full-sky WMAP maps
\cite{bennett/etal:2003}. We shall focus on the V-band ($\sim 61$ GHz) as
it conveniently combines low pixel noise and high spatial resolution,
$21^{\prime}$. In addition, we have also used the W-band and a
foreground ``cleaned'' WMAP map \cite{tegmark/etal:2003} to check that our results
are free from galactic contamination.  We mask out pixels
using the conservative Kp0 mask, that cuts out $21.4 \%$ of the sky
\cite{bennett/etal:2003}. All the maps used have been
digitized into $7^{\prime}$ pixels using HEALPix
\footnote{http://www.eso.org/science/healpix} \cite{gorski/hivon/wandelt:1999}.


\section{Cross-Correlation Estimation}
\label{sec:cross}

We define the cross-correlation function as the expectation value of density
fluctuations $\delta_G= N_G/<N_G>-1$ and temperature anisotropies
$\Delta_T= T-T_0$ (in $\mu$K)
at two positions ${\bf\q_1}$ and ${\bf\q_2}$ in the sky:
$w_{TG}(\theta) \equiv  \langle \Delta_T({\bf\q_1}) \delta_G({\bf\q_2}) \rangle$,
where $\theta = |\bf{\q_2}-\bf{\q_1}|$. 

We compute the CMB-galaxy correlation and the associated statistical
error-bars using the jack-knife (JK) method 
(see \cite{fosalba/gazta:2003} and references therein for further details) 
The survey is divided
into $M$ separate regions the sky, each of equal area
($M=8-16$ depending on the survey used; we find no significant
differences as a function of $M$).
The $w_{TG}$ analysis is then performed
$M$ times, each time removing a different region, the so-called
JK subsamples. The covariance $C_{ij}$ for $w_{TG}$ between
scales $\theta_i$ and $\theta_j$ is obtained by re-scaling the
covariance of the JK subsamples by a factor $M-1$ 
To test the JK errors and their covariance
we have also run $200$ WMAP V-band Monte-Carlo (MC)
realizations of the measured WMAP
temperature angular power-spectrum \cite{bennett/etal:2003} 
and the white noise for the V-band estimated for 1 year of data
\cite{hinshaw/etal:2003}. 

\begin{figure}[t]
\vspace{-3.truecm}
\includegraphics[angle=0,width=0.5\textwidth]{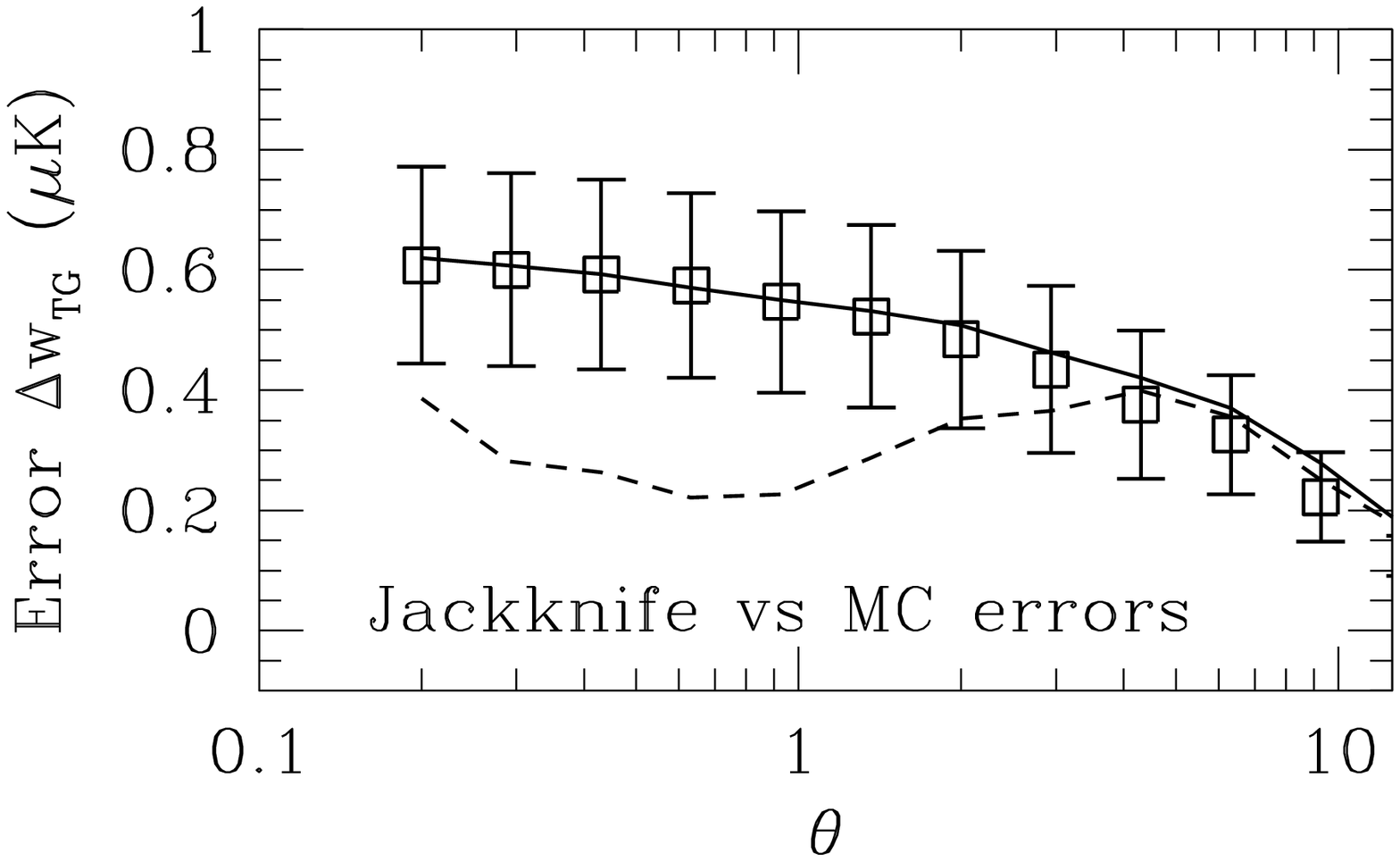}
\includegraphics[angle=0,width=0.5\textwidth]{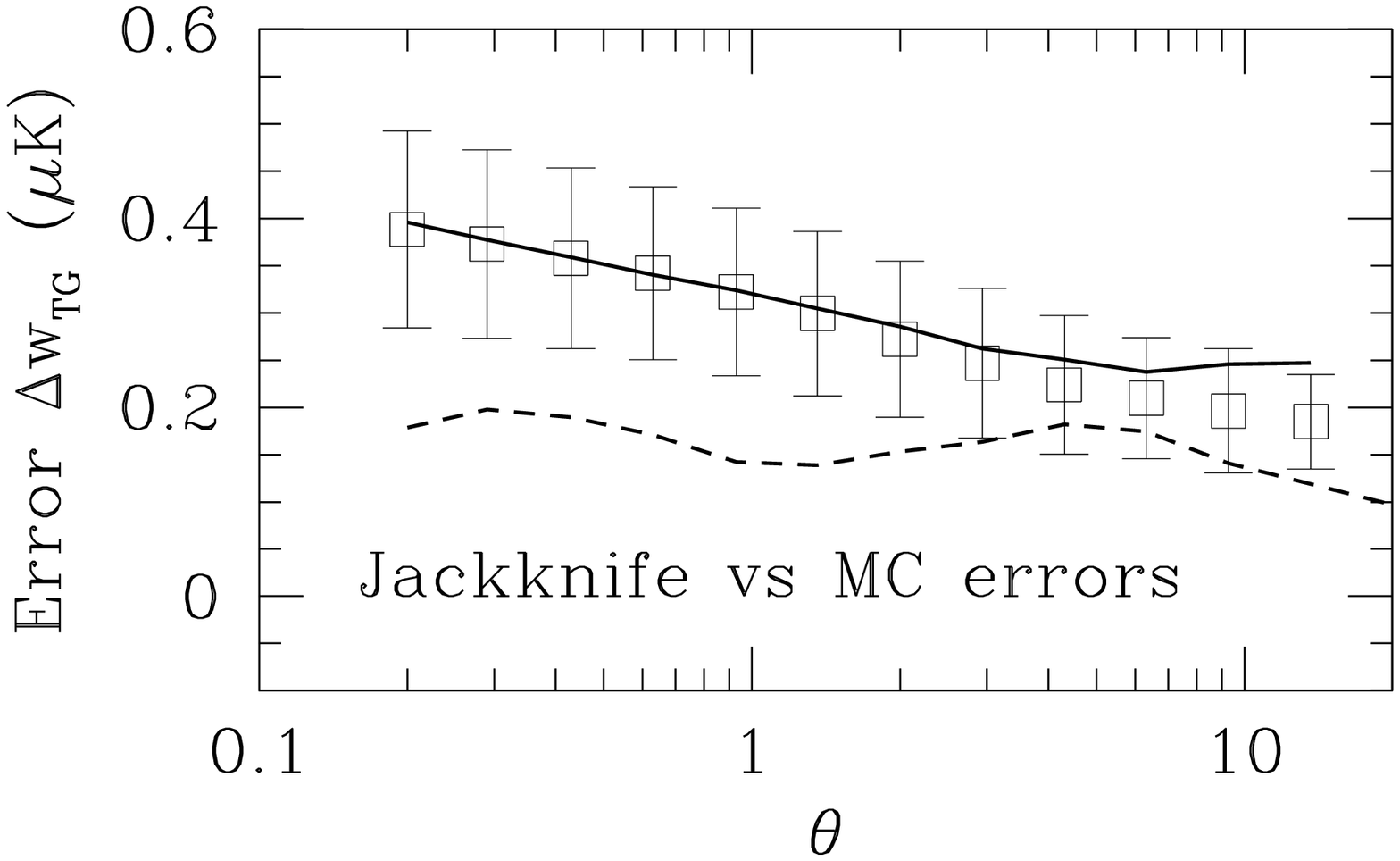}
\caption{{\bf (Left)} error estimation for the WMAP-APM; 
{\bf (Right)} errors for the WMAP-SDSS.
Solid line gives errors from the dispersion in 200
Monte-Carlo simulations; Squares with errobars 
give the mean and dispersion in the jack-knife error estimation 
over the same simulations.
Dashed lines show the jack-knife 
errors in the WMAP-Galaxy correlation for the real samples.
Angular scales are given in degrees.
\label{fig:jkmc}}
\end{figure}

For each MC simulation we estimate the 
mean ''accidental'' correlation $w_{TG}$ of  
simulated CMB maps to the SDSS galaxy density fluctuation map.
We also estimate the associated JK error in each MC simulation.
Fig. \ref{fig:jkmc} compares 
error estimates for the APM (left panel) and SDSS (right) surveys.
They show the 'true' sampling error from the dispersion of
$w_{TG}(\theta)$ in 200 MC simulations, 
along with the mean and dispersion of the JK errors
in the same simulations. The JK and MC errors 
agree within $10-20\%$ for $\theta \simgt 5$ degrees, i.e, the scales
relevant for the ISW effect (see below).
The observed differences in error estimation, particularly on smaller scales, 
are not totally surprising as the MC simulations do
not include any physical correlations
and use a CMB power spectrum that is valid for the
whole sky, and not constraint as to match the CMB power over
the region covered by the galaxy survey. 
Actually, the JK errors provide a model free estimation that
is only subject to moderate ($20\%$) uncertainty, while MC errors
depend crucially on the model assumptions that go into the
simulations.

\begin{figure}[t]
\includegraphics[angle=0,width=0.5\textwidth]{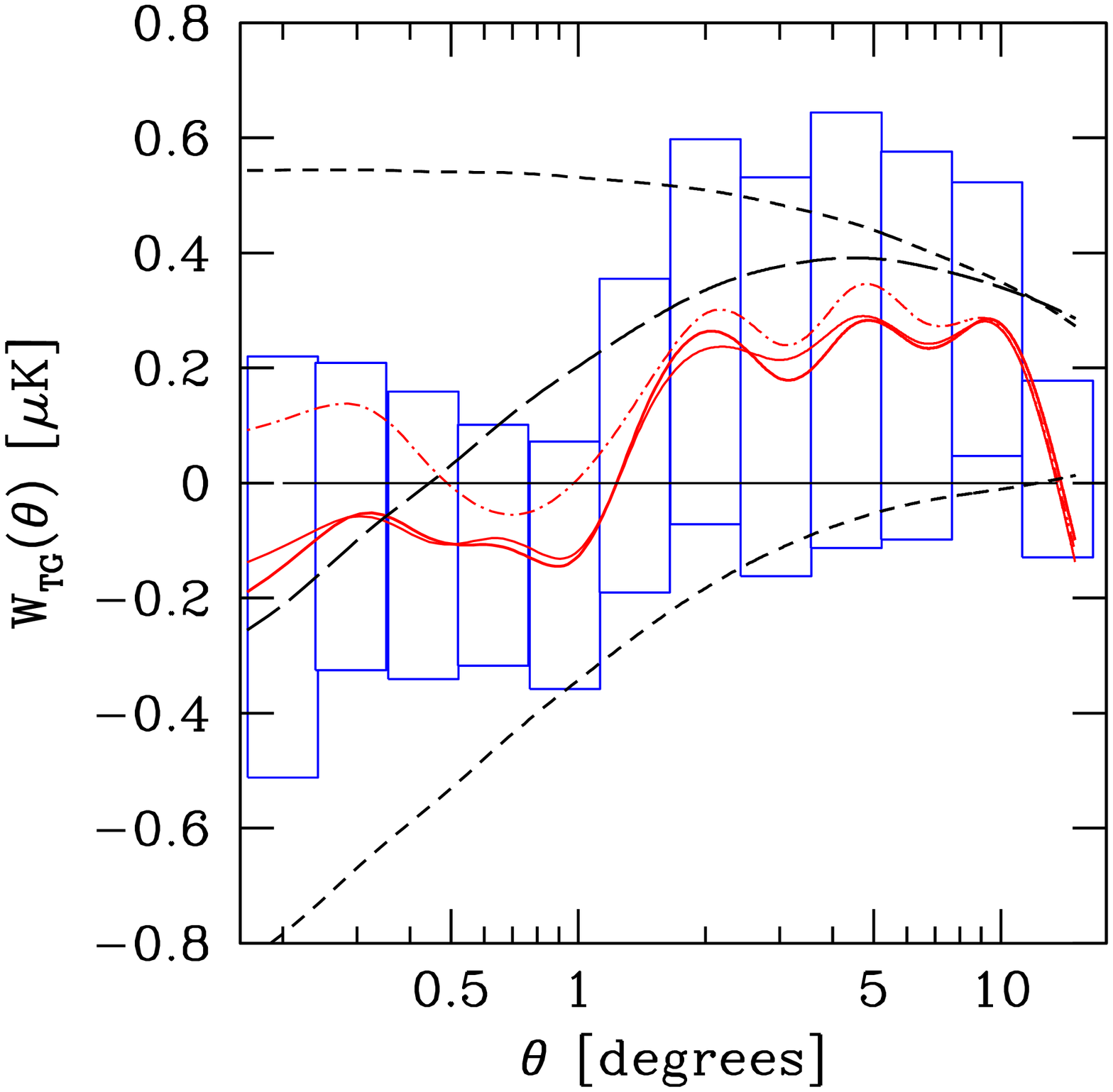}
\includegraphics[angle=0,width=0.5\textwidth]{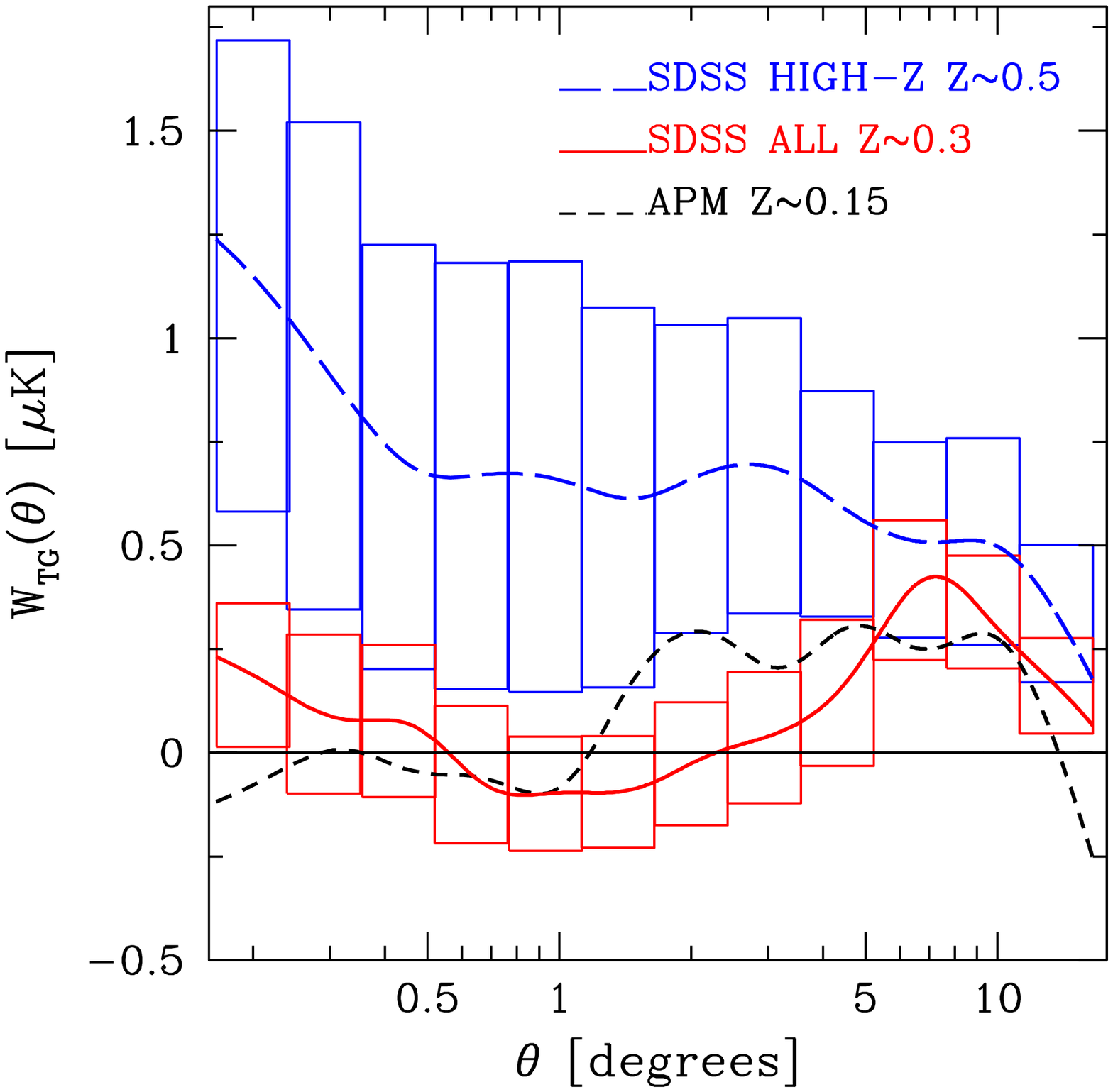}
\caption{ {\bf (Left)} WMAP-APM correlation:
the two solid lines and the dotted
line show $w_{TG}$ results for WMAP bands V, W, and the foreground ``cleaned''
map.  Boxes show the $68\%$ confidence levels.
Also shown are the theoretical predictions 
 for ISW  and SZ (upper and lower short-dashed lines)
and their sum (long-dashed line) for the best-fit \lcdm model.
{\bf (Right)} WMAP-SDSS correlation:
long dashed-line shows the measurement for the {\it SDSS HIGH-Z}
sample, while the solid line displays the correlation for the 
{\it SDSS ALL} sample. For reference, short-dashed line displays the same
measurement using the APM galaxy survey (see left panel) instead of SDSS.
\label{fig:wtg}}
\end{figure}

Fig.\ref{fig:wtg} shows the CMB-galaxy correlation
$w_{TG}$ for the APM (left) and SDSS (right) catalogs
together with the corresponding JK error.
We derive the significance of the detected correlation taking into
account the large (JK)
covariance between neighboring (logarithmic) angular
bins in survey sub-samples.
To assign a conservative
significance for the detection (i.e. against $w_{TG}=0$) we estimate the
minimum $\chi^2$ fit for a constant $w_{TG}$ and give the difference
$\Delta \chi^2 $ to the $w_{TG}=0$ null detection. For example, at
scales  $\theta=4-10^{\circ}$ we get:  
$w_{TG} = 0.35 \pm 0.13 \mu$K for the APM, 
$w_{TG} = 0.26 \pm 0.13 \mu$K for the {\it SDSS all} and 
$w_{TG} = 0.53 \pm 0.21 \mu$K for the {\it SDSS high-z} sample, 
in all cases we give 1-$\sigma$ errorbars.
We find the largest significance in the CMB-galaxy correlation for 
the {\it SDSS high-z} sample: $\Delta \chi^2=9.1$ (ie
probability, $P=0.3\%$ of no detection) for $\theta<10^\circ$.
In order to derive the significance levels for the ISW and SZ effects 
from the observed CMB-Galaxy correlations,
we shall first discuss model predictions.

\section{Comparison to  Theoretical Predictions}
\label{sec:pred}

\begin{figure}[t]
\includegraphics[angle=0,width=0.49\textwidth]{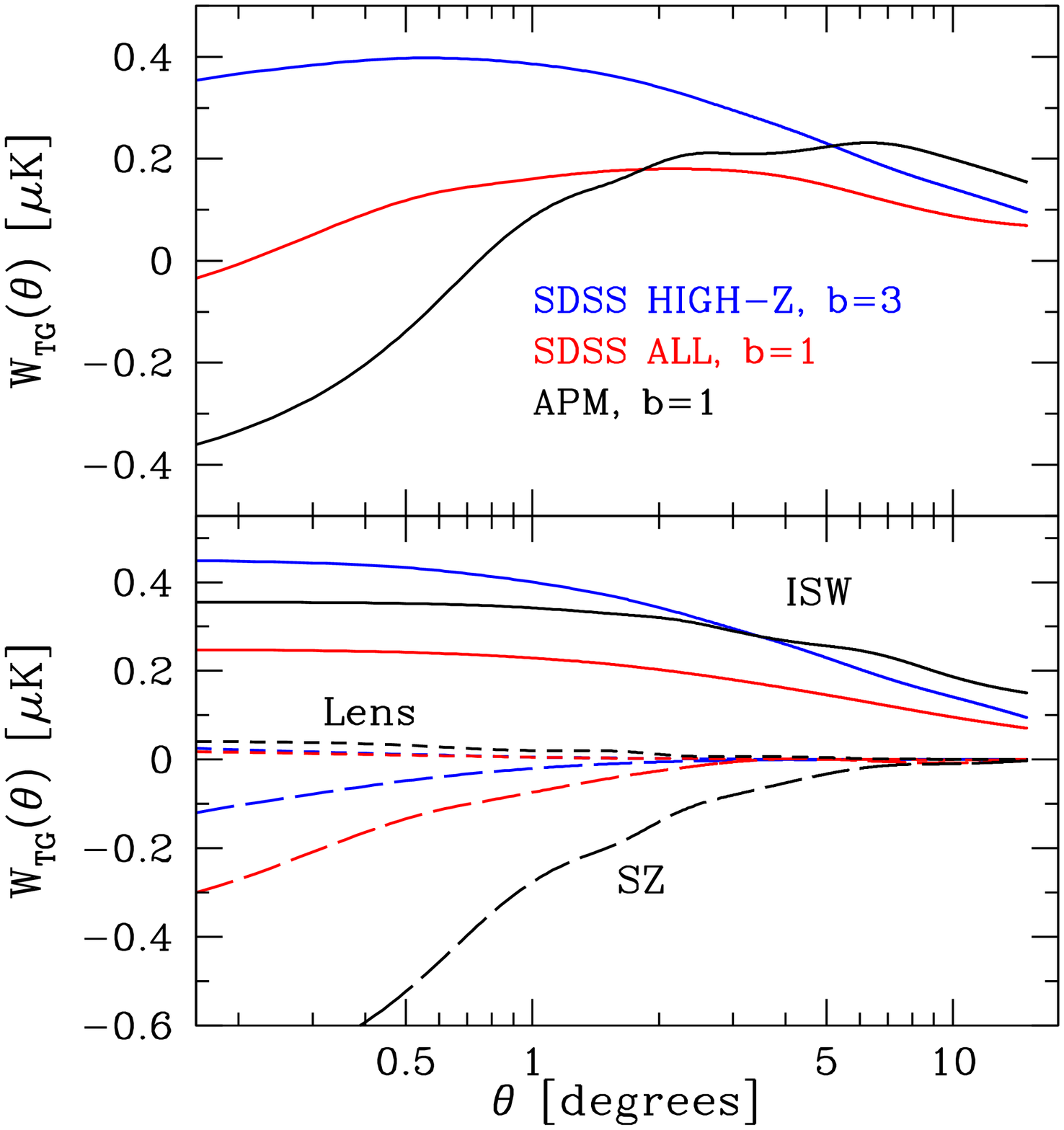}
\includegraphics[angle=0,width=0.51\textwidth]{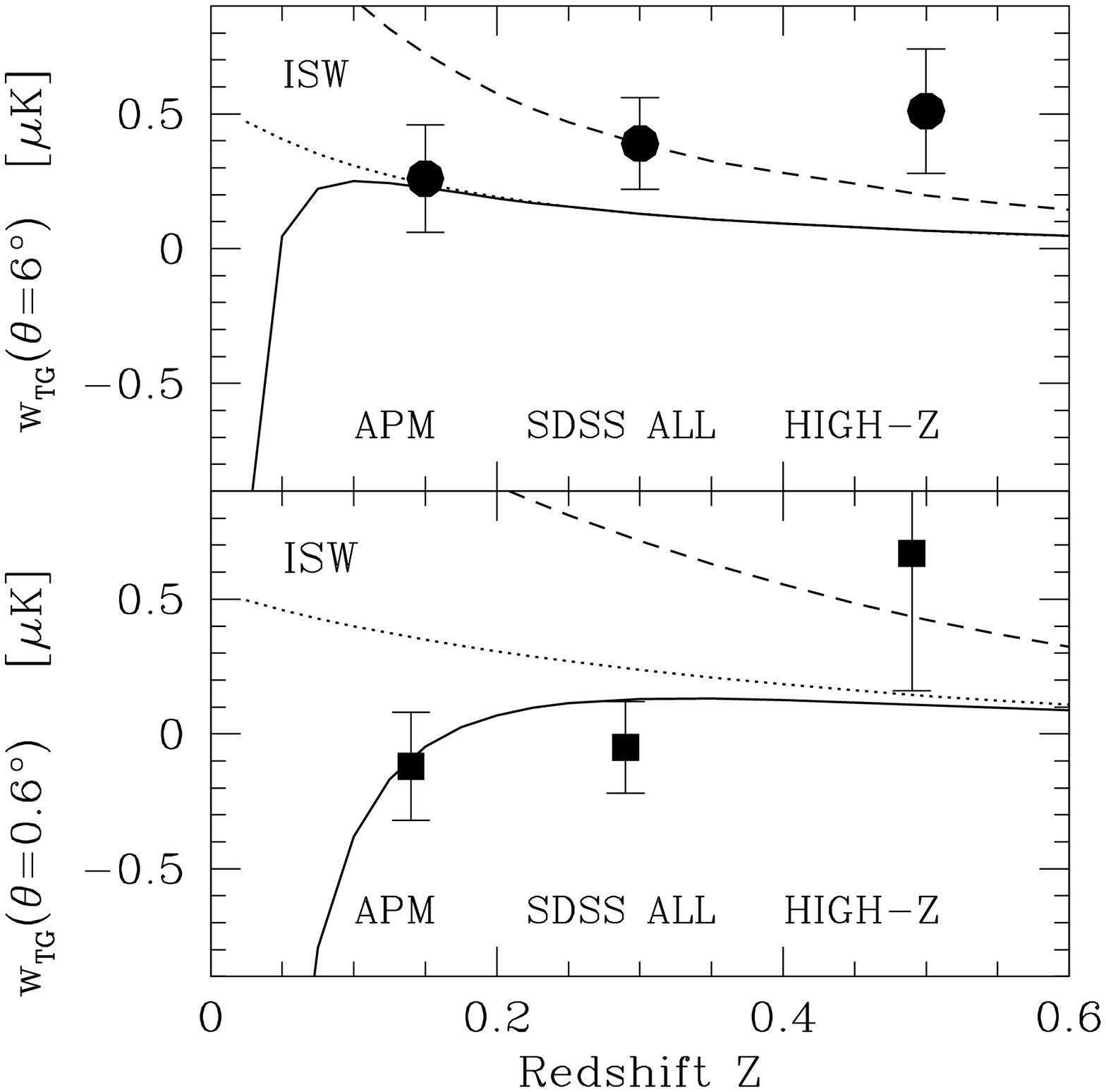}
\caption{
{\bf (Left)} Theoretical predictions: 
(Bottom panel) Continuous, long and short dashed lines show the ISW, SZ
and lensing predictions. Different sets of lines correspond to the
APM (black), {\it SDSS all} (red) and {\it SDSS high-z} (blue) samples. 
(Top panel) The total prediction (ISW+SZ+Lensing)
for the 3 samples. We have assumed a \lcdm model 
with a fixed $b_{gas}=2$ in all cases,
$b=3$ for {\it SDSS high-z} and $b=1$ for APM and {\it SDSS all}.
{\bf (Right)} Predictions and measurements
as a function of redshift: squares and circles correspond
to measured cross-correlation at $\theta=0.6^\circ$ and  $\theta=6^\circ$
respectively. Continuous lines show the corresponding SZ+ISW+lens
predictions for $b=1$ and $b_{gas}=2$. Dotted  and dashed lines shows the ISW 
with $b=1$ and $b=3$ \label{fig:w2pre}.}
\end{figure}

Photons coming to the observer from 
a given direction in the sky ${\bf\q}$ suffer an ISW temperature
change given by: $\Delta T^{ISW}({\bf\q}) = -2 \int dz \dot{\Phi}({\bf\q},z)$, and 
for a flat universe $\nabla^2\Phi = -4\pi
G a^2 \rho_m \delta$. In
Fourier space it reads, $\Phi(k,z) = -3/2 \Omega_m
(H_0/k)^2\delta(k,z)/a$, and thus: 
\be 
w_{TG}^{ISW}(\theta)
= <\Delta_T^{ISW}\delta_G> = \int {dk\over k}~P(k)~g(k\theta) 
\label{eq:wTG_ISW}
\ee 
being, $g(k\theta)={1/{2\pi}} \int dz~W_{ISW}(z)~W_G(z)~j_0(k\theta\,r)$,
where the ISW window function is given by 
$W_{ISW} = -3\Omega_m({H_0/c})^2\dot{F}(z)$, 
with $c/H_0 \simeq 3000 h$ Mpc$^{-1}$,
$\dot{F}=d(D/a)/dr=(H/c)D(f-1)$, and $f \simeq
\Omega_m^{6/11}(z)$ quantifies the time evolution of the gravitational
potential.  The galaxy window function is $W_G \simeq
b(z)~D(z)~\phi_G(z)$, which depends on the galaxy bias, linear
dark-matter growth and the galaxy selection function. The ISW
predictions for the 3 samples are shown in in bottom panel of
Fig.\ref{fig:w2pre}. Unless stated otherwise, we use the concordance
$\Lambda$CDM model with $\Omega_m=0.3$, $\Omega_\Lambda=0.7$, $\Gamma
\simeq h\Omega_m =0.2$ and $\sigma_8=1$.

For the thermal Sunyaev-Zeldovich (SZ) effect, we assume that the gas
pressure $\delta_{gas}$ fluctuations are traced by the galaxy
fluctuations $\delta_{gas} \simeq b_{gas}~\delta_G$ with a relative
amplitude given by the gas bias, $b_{gas} \simeq 2$, representative of
galaxy clusters. A simple estimate of the SZ effect is thus
\cite{refregier/etal:2000},
\be
w_{TG}^{SZ}(\theta) = -b_{gas}~\overline{{\Delta T}} ~w_{GG}(\theta)
\label{eq:sz} 
\ee
where
$\overline{{\Delta T}} = j(x) \overline{y} ~T_0$, 
is the mean temperature change in CMB photons
Compton scattered by electrons in hot intracluster gas,
$T_0 \simeq 2.73$K is the mean CMB temperature, 
$\overline{y} \simeq 1\times 10^{-6}$ is the mean Compton
parameter induced by galaxy clusters estimated for our samples, 
and $j(x) = -4.94$ is the
negative SZ spectral factor for the V-band. 
The weak lensing effect prediction is quite similar to the ISW,
we just need to replace the time derivative of the Newtonian potential by its
2D Laplacian \cite{seljak:1996}:
$W_{Lens} = 3k^2\Omega_m({H_0/c})^2(D/a)/d(r)$,
$d(r)$ being the angular distance to the lensing 
sources.
The total predicted correlation is thus the sum of
three terms: the ISW, thermal SZ and Lensing contributions, 
$w_{TG}=w_{TG}^{ISW}+w_{TG}^{SZ}+w_{TG}^{Lens}$.
Fig.\ref{fig:w2pre} shows individual contributions of these effects
(bottom panel) and the total (top) for the 3 samples analyzed. 
The ISW effect typically dominates for angles
$\theta >4^{\circ}$, while the SZ effect is expected to be significant
on small scales $\theta<1^{\circ}$. Lensing is found to be negligible
at all scales for our samples.

Before one can make a direct comparison between theory and observations,
the issue of galaxy bias must be addressed.
As shown in Fig \ref{fig:w2pre} (see plot on the right), the higher (SDSS) 
redshift sample
requires a high bias $b > 1$ to explain the large cross-correlation
seen at all scales. At low redshifts 
the measured correlation is dominated by the thermal
SZ on small scales $\theta<1^\circ$, while the ISW 
still dominates on large-scales, $\theta>4^\circ$. Here no bias is required to
reproduce the observations. 
Actually, this agrees quite well with our self-consistent bias
estimation: for each sample we estimate the ratio 
$b^2 = w_{GG}/w_{MM}$, where
$w_{MM}$ and $w_{GG}$ are the (theoretically predicted)
matter and (measured) galaxy auto-correlation 
functions. For APM and {\it SDSS all} samples we find $b^2 \simeq 1$,
while for the {\it SDSS high-z} sample we get $b^2 \simeq 6$.

\subsection{Significance Tests}
\label{sec:chi2}

\label{sec:isw}

\underline{\bf ISW effect:}
On large scales $\theta>4^\circ$, the ISW effect is expected to
dominate for all survey depths (see Fig \ref{fig:w2pre}).  Therefore,
from the large-angle CMB-Galaxy correlation, we can directly infer the
ISW effect (ie, $w_{TG}=w^{ISW}_{TG}$, see end of \S\ref{sec:cross}).  
In particular, for the APM survey, a
constant correlation fit rejects the null detection with high
significance $\Delta \chi^2=6.0$ ($P=1.4\%$), similar to
the values for the {\it SDSS high-z} sample $\Delta \chi^2=6.1 ~(P=1.3\%)$.
A smaller significance is obtained from the {\it SDSS all} sources: 
$\Delta \chi^2 =3.9$ (P$=4.8\%$). 
Since these samples are basically independent, we can combine them to
infer a total significance for the ISW detection: we find a total
$\Delta \chi^2=16$ ($P=0.1 \%$ for 3 d.o.f) corresponding to a 3.3$\sigma$.
Note we could do better using a (scale dependent) \lcdm model
theory prediction, but at the cost of introducing model dependent
detection levels.  Moreover, we can further include the ISW-dominated
small angle bins in our deepest sample, where SZ is negligible,
increasing the significance to $\Delta \chi^2=18.8$, 
($P=0.03\%$ for 3 d.of.), ie
we detect the ISW effect at the a 3.6$\sigma$ level.

\label{sec:sz}

\underline{\bf SZ effect:}
We can estimate the significance of the drop in the signal at small
angles in the APM and {\it SDSS all} samples due to the SZ effect 
(see Fig\ref{fig:w2pre})
using the best-fit constant at large angles (ie, the ISW signal) 
and ask for the observed
deviation from such value at smaller scales. For $\theta<1^{\circ}$, we find:
$w^{SZ}_{TG}=-0.41\pm 0.16$ for APM, and
$w^{SZ}_{TG}=-0.27\pm 0.11$ for {\it SDSS all} (1-$\sigma$ errorbars).
This test gives $\Delta \chi^2 =8.5 ~(P=0.3\%)$ for the APM sample
and $\Delta \chi^2 =5.5 ~(P=2\%)$ for the {\it SDSS all}.

\section{Discussion}
\label{sec:discuss}

\begin{figure}
\centering{
\epsfysize=6.5cm \epsfbox{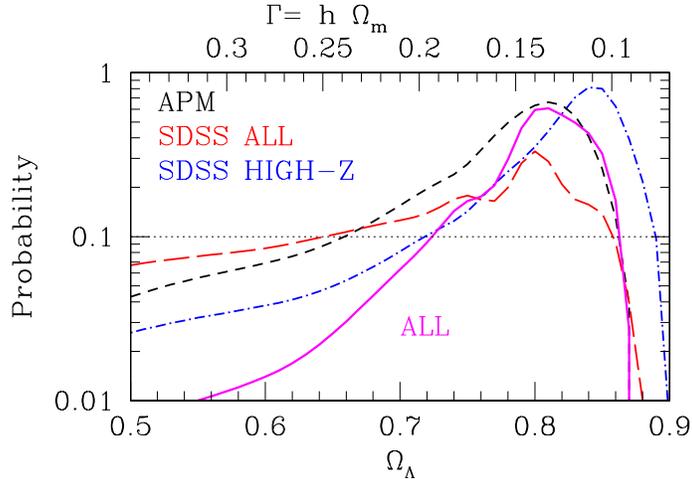}}
\caption{\label{fig:FitLambda}
Estimating dark-energy: Long-dashed, short-dashed and dot-dashed lines show the
probability distribution for $\Omega_\Lambda$ in the {\it SDSS all}, APM and
{\it SDSS high-z} samples. Solid line shows the combined distribution.}
\end{figure}

We have cross-correlated WMAP with
the APM and SDSS DR1 optical galaxy surveys. Our analysis 
includes $5800$ deg$^2$ (14 $\%$ of the sky) and comprises
$7.6$ million galaxies in compact regions of the
north and south hemispheres. We obtain significant
cross-correlations for galaxy samples in a wide redshift range 
(z $\sim 0.15-0.5$).
We detect a positive large-scale correlation which is in good 
agreement with the ISW effect and its redshift evolution 
as expected from \lcdm models.
The combined analysis for the 3 samples investigated (APM, SDSS-all and SDSS high-z)
yields a $99.97 \%$ ISW detection level (3.6$\sigma$).

Before using matter predictions to estimate
parameters from the observed CMB-galaxy correlation, 
we self-consistently estimated the galaxy bias $b$ by comparing 
the matter angular auto-correlation function in
each model to the measured galaxy auto-correlation in each sample.  

Once galaxy bias is determined, we find that all the 
measured cross-correlations on large scales are in good agreement with ISW
predictions for a dark-energy dominated universe.  Fig \ref{fig:FitLambda} shows the
probability distribution for $\Omega_\Lambda$ in a flat \lcdm model. We
have fixed $\sigma_8=1$, $h=0.7$ and $\Omega_M+\Omega_\Lambda=1$.  As
we vary $\Omega_\Lambda$ the shape parameter for the linear power
spectrum $P(k)$ consistently changes $\Gamma = h \Omega_M$. We only use the data for
$\theta>4^\circ$, where the ISW is the dominant contribution for all samples. 

From our analysis a coherent dark-energy dominated universe arises:
both APM and SDSS samples point to large values of $\Omega_\Lambda$, 
with the best fit $\Omega_\Lambda \simeq 0.8$ and a rather narrow 2$\sigma$ range
$\Omega_\Lambda=0.69-0.87$. We stress that this dark-energy estimation
is independent from other known probes (e.g, SN type Ia data).

We have also found evidence for the thermal SZ effect
from the drop of the CMB-galaxy correlation on small-scales in the
low-z samples of APM and SDSS galaxies. The estimated SZ effect is
compatible with a Compton parameter $\overline{y} \simeq 1\times 10^{-6}$.
These new measurements can be used to constrain the redshift evolution of the
physical properties of gas inside galaxy clusters.

Upcoming wider and deeper Galaxy surveys
(e.g future data releases from the SDSS) in combination with higher-spatial resolution
and sensitivity full-sky CMB maps (e.g. 2yr-WMAP data, PLANCK) 
shall eventually allow for a higher significance detection of both
the ISW and SZ signals. Moreover these new generation surveys
can potentially extract the challenging lensing signal on large-scales
\cite{seljak:1996,hu:2002,kesden/etal:2003}.
This program will eventually provide us with a better understanding
of the gravitational instability picture and 
set new and tighter constraints on basic cosmological parameters.

\section*{Acknowledgments}
We acknowledge support from the Barcelona-Paris bilateral project
(Picasso Programme).  PF acknowledges support from a post-doctoral CMBNet
fellowship from the EC and grant NASA ATP NAG5-12101.  
EG and FJC acknowledge the Spanish MCyT, project AYA2002-00850,
EC-FEDER funding.

\end{document}